\documentclass[aps,prb,twocolumn,showpacs,showkeys,superscriptaddress]{revtex4-1}
\newif\ifshowdetails
\showdetailstrue

\usepackage{graphics}
\usepackage{color}
\usepackage{amsmath}
\usepackage{bm}

\input{MyMc}

\newcommand{\bo}{\ensuremath{\beta_{1}}}
\newcommand{\bp}{\ensuremath{\beta_{2}}}
\newcommand{\bsum}{\ensuremath{\beta_{\Sigma}}}
\newcommand{\bboth}{\ensuremath{\beta_{1,2}}}

\begin{document}

\title{%
  Tunable $\pm\varphi$, $\varphi_0$ and $\varphi_0\pm\varphi$ Josephson junction
}

\author{E. Goldobin}
\author{D. Koelle}
\author{R. Kleiner}
\affiliation{%
  Physikalisches Institut and Center for Collective Quantum Phenomena in LISA$^+$,
  Universit\"at T\"ubingen, Auf der Morgenstelle 14, D-72076 T\"ubingen, Germany
}

\date{%
  \today
}

\begin{abstract}
  We study a 0-$\pi$ dc superconducting quantum interference device (SQUID) with asymmetric inductances and critical currents of the two Josephson junctions (JJs). By considering such a dc SQUID as a black box with two terminals, we calculate its effective current-phase relation $I_s(\psi)$ and the Josephson energy $U(\psi)$, where $\psi$ is the Josephson phase across the terminals. We show that there is a domain of parameters where the black box has the properties of a $\varphi$ JJ with degenerate ground state phases $\psi=\pm\varphi$. The $\varphi$ domain is rather large, so one can easily construct a $\varphi$ JJ experimentally. We derive the current phase relation and show that it can be tuned \emph{in situ} by applying an external magnetic flux resulting in a continuous transition between the systems with static solutions $\psi=\pm\varphi$, $\psi=\varphi_0$ ($\varphi_0 \neq 0,\pi$) and even $\psi=\varphi_0\pm\varphi$. The dependence of $\varphi_0$ on applied magnetic flux is not $2\pi$ (one flux quantum) periodic.

\end{abstract}



\maketitle

\section{Introduction}

One of the most important characteristics of a Josephson junction (JJ) is its current-phase relation (CPR), \ie, the  relation between the supercurrent $I_s$ flowing through the junction and the Josephson phase $\phi$ across it. The CPR, aka the first Josephson relation, plays a fundamental role and is responsible for almost all properties of the JJ. The CPR usually depends on the microscopic physics of the Josephson barrier. The CPR is usually a $2\pi$ periodic function of $\phi$ and, in the most simple case, is given by $I_s=I_c\sin\phi$. A review of different types of CPRs can be found elsewhere\cite{Golubov:2004:CPR}.

The quantity directly related to the CPR is the Josephson energy profile $U(\phi)$. It is defined so that $I_s(\phi)=\divp{U(\phi)}{\phi}$. For a sinusoidal CPR $U(\phi)=E_J(1-\cos\phi)$, where $E_J=\Phi_0 I_c/2\pi$ is the Josephson energy, and $\Phi_0\approx 2.068\units{fWb}$ is the magnetic flux quantum. $U(\phi)$ is also a $2\pi$ periodic function of $\phi$.

The most interesting JJs both in terms of applications and fundamental studies are those having a non-trivial Josephson energy profile $U(\phi)$. For example, $\pi$ JJs that received a lot of attention\cite{Bulaevskii:pi-loop,Buzdin:SfiS:pi-JJ,Baselmans:1999:SNS-pi-JJ,Kontos:2002:SIFS-PiJJ,Oboznov:2006:SFS-Ic(dF),Weides:2006:SIFS-HiJcPiJJ,vanDam:2006:QuDot:SuperCurrRev,Jorgensen:2007:QuDotJJ:0-pi-transition,Feofanov:2010:SFS:pi-qubit,Khabipov:2010:RSFQ:pi-flip-flop,Vasenko:2011:SIFS.IVC} have negative critical current, which results in the sign change of $U(\phi)$. As a result, the energy minimum (and the ground state) corresponds to $\phi=\pi$, where the usual JJ (called 0 JJ in this context) has a maximum. Further, there was a theoretical predictions\cite{Buzdin:2008:varphi0,Alidoust:2013:Geom-varphi0} that one can obtain the so-called $\varphi_0$ JJ, \ie, the one having a single $U(\phi)$ minimum (with each $2\pi$ period) situated at $\phi=\varphi_0\neq0,\pi$.  More recently, $\varphi$ JJs having a periodic double-well potential $U(\phi)$ and, therefore, a degenerate ground state (two $U(\phi)$ minima) with the phases $\phi=\pm\varphi$ were investigated\cite{Goldobin:2011:0-pi:H-tunable-CPR,Sickinger:2012:varphiExp,Goldobin:2013:varphi-bit,Lipman:2014:varphiEx}. Upon application of magnetic field one can also obtain a $U(\phi)$ without reflection symmetry.

Theoretically, it was predicted\cite{Tanaka:1996:d-wave-JJ.T,Tanaka:1997:AnisoSC.JJ} that a JJ made of $d$-wave superconductors in a specific range of parameters (orientation angle, temperature, \etc) can posses a $\varphi$ ground state. A degenerate ground state was obtained from measurements of the CPR of \textit{d}-wave based nano JJs\cite{Ilichev:2001:Sym45GBJJ:DegenGrdStt}. Later on, some other indications of a $\varphi$ state, such as an anomalous temperature dependence of the critical current, were observed\cite{Testa:2005:TiltGB:NanoJJ.Transport} also on nano JJs. The faceting along longer grain-boundary JJs based on \textit{d}-wave superconductors results in an effective (facet-averaged) $\varphi$ JJ\cite{Mints:1998:SelfGenFlux@AltJc}. In the latter case one expects non-quantized splintered vortices\cite{Mints:2001:FracVortices@GB,Goldobin:CPR:2ndHarm}, which were observed\cite{Mints:2002:SplinteredVortices@GB} using superconducting quantum interference device (SQUID) microscopy. However, in all cases no state manipulation or readout were demonstrated, probably because of high damping and poor control over JJ properties.

Recently, we have suggested\cite{Goldobin:2011:0-pi:H-tunable-CPR} and successfully demonstrated\cite{Sickinger:2012:varphiExp} a $\varphi$ JJ based on conventional low-$T_c$ superconductors with tailored ferromagnetic barrier\cite{Weides:2007:Ferro-0-pi-JJ,Kemmler:2010:SIFS-0-pi:Ic(H)-asymm}. This junction has a degenerate ground state phase $\pm\varphi$, i.e. its Josephson energy profile looks like a $2\pi$-periodic double well potential (in the absence of bias current). The two ground states can be used to store information\cite{Goldobin:2013:varphi-bit}. The unusual physics of $\varphi$ JJs was discussed in several works\cite{Goldobin:CPR:2ndHarm,Goldobin:2013:RetrapButfly}. However, the $\varphi$ JJs constructed so far are rather large. Making them smaller (shorter) requires very exact control on parameters such as critical current densities and the lengths of the 0 and $\pi$ regions.

In this paper, we propose an effective $\varphi$ JJ based on an asymmetric 0-$\pi$ dc SQUID, \ie, a dc SQUID with one 0 and one $\pi$ JJ, with finite inductance and asymmetric critical currents of the 0 and $\pi$ JJ. This not only has advantages over the previous proposals\cite{Goldobin:2011:0-pi:H-tunable-CPR,Sickinger:2012:varphiExp,Lipman:2014:varphiEx} in terms of geometrical dimensions, margins, and the size of the $\varphi$ domain in parameter space, but also shows other unique features. For example, it can be operated not only as a $\varphi$ JJ, but also as a $\varphi_0$ JJ\cite{Buzdin:2008:varphi0} or as a combination of both, \ie, as a $\varphi_0\pm\varphi$ JJ. This goes much beyond the trivial use of a dc SQUID as the substitution for a single JJ with magnetic-field-tunable critical current.

The paper is organized as follows. In sec.~\ref{Sec:Model} we introduce the considered model system and the equations describing it. In the main sec.~\ref{Sec:Results} we present our numerical results. Sec.~\ref{Sec:Conclusions} summarizes our findings. In appendix \ref{Sec:GL} we consider the vicinity of the $0\leftrightarrow\varphi$ transition in parameter space and derive many results analytically similar to the Ginzburg-Landau approach.

\section{Model}
\label{Sec:Model}

The SQUID circuit is shown in the inset of Fig.~\ref{Fig:CPR}. The applied bias current $I$ splits into two branches with inductances $L_1$ and $L_2$ and Josephson junctions with critical currents $I_{c1}$ and $I_{c2}$.

For the sake of simplicity, we will derive everything in normalized units. The current will be normalized to the largest (by absolute value) critical current of the two JJs. Without loss of generality we assume that $I_{c1}>0$ and $|I_{c2}| \leq I_{c1}$. Then we define $\alpha=I_{c2}/I_{c1}$ as the normalized critical current of the second JJ, with $|\alpha|\leq1$. Thus, $\alpha>0$ corresponds to a 0-0 SQUID, while $\alpha<0$ to a 0-$\pi$ SQUID. Although, our main focus is on a 0-$\pi$ SQUID ($\alpha<0$) the results for a 0-0 SQUID will be included automatically as well. Further, we introduce normalized inductances
\begin{equation}
  \bo = \frac{2\pi I_{c1}L_1}{\Phi_0}, \quad \bp = \frac{2\pi I_{c1}L_2}{\Phi_0}
  . \label{Eq:betas.def}
\end{equation}
Note that the definition of $\bp$ uses $I_{c1}$, so that by changing $\bo$ and $\bp$ one can see the effect of inductances only, while changing $\alpha$ one can see the effect of critical current asymmetry only. These definitions are related to the conventional\cite{Clarke:2006:SQUID-book}
$\beta_L=2I_{c1}L_\Sigma/\Phi_0$, as $\pi\beta_L=\bo+\bp=\bsum$, where $L_\Sigma=L_1+L_2$.

The total phase $\psi$ across the SQUID, see the inset of Fig.~\ref{Fig:CPR}, can be expressed in two ways
\begin{subequations}
  \begin{eqnarray}
    \psi &=& \phi_1 + \phantom{\alpha}\bo\sin\phi_1 + r_1\phi_{e}
    ; \label{Eq:phi_1}\\
    \psi &=& \phi_2 + \alpha          \bp\sin\phi_2 - r_2\phi_{e}
    , \label{Eq:phi_2}
  \end{eqnarray}
  \label{Eq:phi_12(psi)}
\end{subequations}
where $r_1+r_2=1$, and $r_1$ and $r_2$ are the ratios, in which the externally applied normalized flux/phase $\phi_e=2\pi f=2\pi\Phi_e/\Phi_0$ ($f$ is the normalized flux, aka frustration) is divided between the two branches. If the external magnetic field is applied by a coil, $r_{1,2}\phi_{e}=2\pi M_{1,2} I_0/\Phi_0$, where $I_0$ is the current in the coil creating the magnetic field and $M_{1,2}$ is the mutual inductance between this coil and the left and the right arms of the SQUID, respectively. For the sake of simplicity, we neglect the mutual inductance between $L_1$ and $L_2$.

The sum of the currents in both branches is given by
\begin{equation}
  \gamma = \sin\phi_1 + \alpha \sin\phi_2
  , \label{Eq:gamma}
\end{equation}
where $\gamma=I/I_{c1}$ is the normalized bias current.

Since we are interested in a geometrically small system which should not have too many internal states, we focus on the case of small, but finite, inductances, i.e., $0\leq\bo\leq 1$ and $0\leq\bp\leq1$.

\section{Results}
\label{Sec:Results}

The easiest way to solve Eqs.~\eqref{Eq:phi_12(psi)} and \eqref{Eq:gamma} is to calculate $\phi_1$ and $\phi_2$ from Eqs.~\eqref{Eq:phi_12(psi)} for given $\psi$. Note that for $\bo,\bp,|\alpha|\leq1$ each of Eqs.~\eqref{Eq:phi_12(psi)} has a \emph{unique} solution $\phi_1$ or $\phi_2$ for given $\psi$. It is convenient to define a universal function $\phi(v,p)$, which is a solution $\phi$ of the equation $v=\phi+p\sin(\phi)$ for given argument $v$ and parameter $|p|\leq 1$. Then $\phi_1(\psi)=\phi(\psi-r_1\phi_e,\bo)$ and $\phi_2(\psi)=\phi(\psi+r_2\phi_e,\alpha\bp)$. The function $\phi(v,p)$ has an obvious, but useful, property:
\begin{eqnarray}
  \phi(v+\pi n,p) &=& \pi n +\phi[v,(-1)^n p]
  , \label{Eq:fnphi.prop}
\end{eqnarray}
where $n$ is an integer.

\subsection{Current-phase relation and Josephson energy}

\begin{figure}[!tb]
  \includegraphics{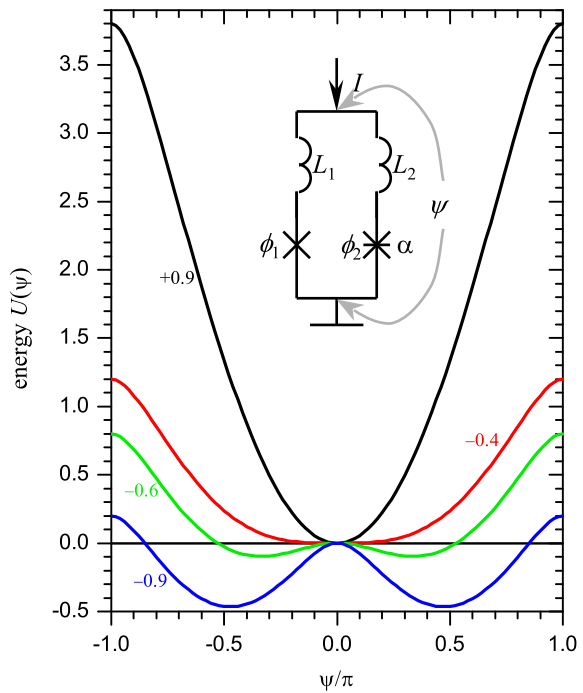}
  \caption{(Color online)
    The energy $U(\psi)$ of the system given by Eq.~\eqref{Eq:U(psi)} for different values of the asymmetry parameter $\alpha$ and for $\bo=\bp=0.7$ ($\alpha_c\approx-0.417$).
    Inset shows the schematics of the circuit considered.
  }
  \label{Fig:CPR}
\end{figure}

To determine the CPR $\gamma(\psi)$ we calculate $\phi_1(\psi)$ and $\phi_2(\psi)$ as mentioned above and then use Eq.~\eqref{Eq:gamma} to obtain $\gamma$. We do not show $\gamma(\psi)$ plots here, but show Josephson energy $U(\psi)$ plots instead.

The total energy of the system is given by
\begin{equation}
  U(\psi)=U_J(\psi)+U_L(\psi)
  , \label{Eq:U(psi)}
\end{equation}
where
\begin{equation}
  U_J(\psi) = [1-\cos\phi_1(\psi)] + \alpha[1-\cos\phi_2(\psi)]
  , \label{Eq:U_J(psi)}
\end{equation}
is the Josephson energy of both JJs, and
\begin{equation}
  U_L(\psi) = \frac{\bo}{2}\sin^2\phi_1(\psi) + \frac{\bp}{2}\alpha^2\sin^2\phi_2(\psi)
  , \label{Eq:U_L(psi)}
\end{equation}
is the magnetic field energy stored in the inductors. By direct substitution, one can see that $U'(\psi)\equiv\gamma(\psi)$ (here and below the prime denotes $\divp{}{\psi}$ by default), like for any JJ.

Consider the case of zero applied magnetic flux $\phi_e=0$. Several examples of $U(\psi)$ are presented in Fig.~\ref{Fig:CPR}. At positive $\alpha$ (conventional 0-0 SQUID) the energy profile resembles the usual $U_\mathrm{conv}(\psi)=1-\cos(\psi)$ profile of a conventional single JJ with the ground state at $\psi=0$. Deviations from $U_\mathrm{conv}(\psi)$ are due to finite inductances and make $U(\psi)$ sharper than $U_\mathrm{conv}(\psi)$ near the maxima and more shallow than $U_\mathrm{conv}(\psi)$ near the minima. As $\alpha$ decreases down to 0, the height of $U(\psi)$ decreases as $2(1+\alpha)$ while the shape of $U(\psi)$ almost does not change. As $\alpha$ becomes negative (0-$\pi$ SQUID) the $U(\psi)$ becomes more flat near $\psi=0$. For $\alpha$ below some critical value $\alpha_c$ the energy profile develops two minima, as can be seen in Fig.~\ref{Fig:CPR}. The system has a degenerate ground state $\psi=\pm\varphi$, \ie, forms a $\varphi$ JJ \cite{Tanaka:1996:d-wave-JJ.T,Tanaka:1997:AnisoSC.JJ,Mints:1998:SelfGenFlux@AltJc,Buzdin:2003:phi-LJJ,Goldobin:CPR:2ndHarm,Goldobin:2011:0-pi:H-tunable-CPR,Lipman:2014:varphiEx,Sickinger:2012:varphiExp}.

The transition to the degenerate ground state takes place at the value of $\alpha=\alpha_c$, for which $U''(0)=\gamma'(0)=0$. This allows to calculate
\begin{equation}
  \alpha_c = \frac{-1}{1+\bo+\bp} = \frac{-1}{1+\bsum} = \frac{-1}{1+\pi\beta_L}
  . \label{Eq:alpha_c}
\end{equation}
This is one of the central results of the paper.

\begin{figure}[!tb]
  \includegraphics{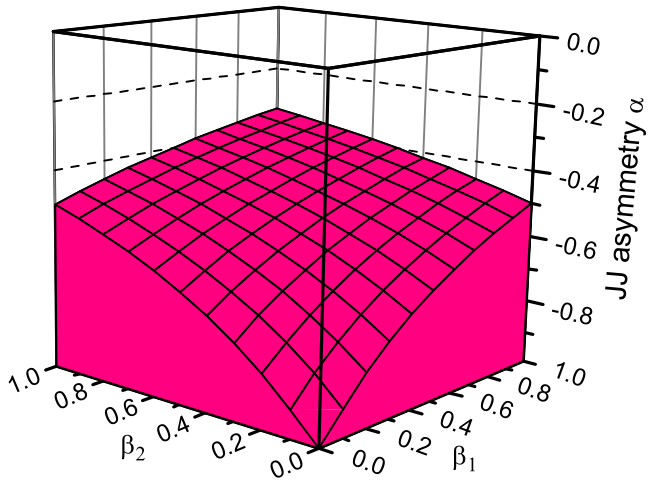}
  \caption{(Color online)
    The domain of the $\varphi$ state (pink/gray) as a function of parameters $\alpha$, $\bo$ and $\bp$.
  }
  \label{Fig:VarphiDomain.3D}
\end{figure}

The domain of the $\varphi$ state is shown in Fig.~\ref{Fig:VarphiDomain.3D}. From the sides it is limited by our choice of parameters $0<\bboth<1$ and from the bottom by $\alpha>-1$. From the top it is limited by the surface $\alpha_c(\bo,\bp)$ given by Eq.~\eqref{Eq:alpha_c}. The $\varphi$ domain has the maximum size (height) along the $\alpha$ axis equal to $-1 \leq \alpha \leq -1/3$ for $\bboth=1$. For $\bboth\to0$ the height of the $\varphi$ domain along the $\alpha$ axis vanishes linearly $\propto \bsum$. We also note that in the case of $\bsum=0$, used by many authors because it is solvable analytically, one \emph{never} obtains a $\varphi$ JJ.

An important practical difference between the 0-$\pi$ dc SQUID considered here and a 0-$\pi$ JJ of finite length\cite{Mints:1998:SelfGenFlux@AltJc,Buzdin:2003:phi-LJJ,Goldobin:2011:0-pi:H-tunable-CPR,Sickinger:2012:varphiExp} is that our system has a rather large $\varphi$ domain where the theory presented here works. Even in the case of $\bsum\to0$ the range of $\alpha=-1\ldots\alpha_c$, corresponding to the $\varphi$ domain, shrinks as $\Delta\alpha=\alpha_c-(-1)=\bsum$, \ie, linearly. Instead, for 0-$\pi$ JJs the $\varphi$ domain shrinks as\cite{Buzdin:2003:phi-LJJ,Goldobin:2011:0-pi:H-tunable-CPR} $\delta\approx\frac13\mean{L}^3$, where $\delta=(L_0-L_\pi)/2$ is the deviation of the 0 and $\pi$ facet length from the average length $\mean{L}=(L_0+L_\pi)/2$. This imposes additional requirements on fabrication accuracy or one should move to a region of the phase diagram where the theory works only qualitatively, as in the first experiments\cite{Sickinger:2012:varphiExp,Goldobin:2013:varphi-bit}. In other works\cite{Pugach:2010:CleanSFS:varphi-JJ,Heim:2013:varphi-JJ/S-FN-S} some alternative techniques to enlarge $\varphi$ domains were suggested. However, the relative volume of the $\varphi$ domains in parameter space is still much smaller than in the present work.


Let us now consider $U(\psi)$ at finite applied magnetic flux $\phi_e\neq0$. First, we point out that Eqs.~\eqref{Eq:phi_12(psi)} and \eqref{Eq:gamma} have several important symmetry properties, when the applied flux changes by a half-integer number of flux-quanta $\Phi_0$
\begin{subequations}
  \begin{eqnarray}
    \phi_e^\mathrm{new} &=& \phi_e+n\pi;\\
    \psi  ^\mathrm{new} &=& r_1 n\pi;\\
    \phi_1^\mathrm{new} &=& \phi_1;\\
    \phi_2^\mathrm{new} &=& \phi_2+n\pi;\\
    \alpha^\mathrm{new} &=& (-1)^n \alpha;\\
  \end{eqnarray}
  \label{Eq:symmetry.nPi}
\end{subequations}
Upon such transformation the current does not change, \ie,
\begin{equation}
  \gamma(\psi,\alpha) = \gamma(\psi^\mathrm{new},\alpha^\mathrm{new})
  , \label{Eq:CPR.new}
\end{equation}
while the Josephson energy, which is still $2\pi$ periodic in $\psi$, can be expressed in terms of new variables as
\begin{equation}
  U(\psi,\alpha) = U(\psi^\mathrm{new},\alpha^\mathrm{new})+\alpha^\mathrm{new}[(-1)^n-1]
  . \label{Eq:U.new}
\end{equation}
Thus, by applying a half-integer $f$ we can turn a 0-0 SQUID with $\alpha>|\alpha_c|$ (effective 0 JJ) into a 0-$\pi$ SQUID (effective $\varphi$ JJ) with $\alpha^\mathrm{new}=-\alpha<\alpha_c$ and vice versa. An even more interesting point is that $\psi$ shifts by an amount $r_1 n\pi$, which is, in general, not a multiple of $2\pi$. Since after transformation \eqref{Eq:symmetry.nPi} the value of $\gamma$ does not change (it depends only on $(\phi_1,\phi_2) \bmod 2\pi$), the CPR $\gamma(\psi)$ shifts along the $\psi$ axis by the amount $r_1 n\pi$. This is another central result of the paper.

\begin{figure}[!htb]
  \includegraphics{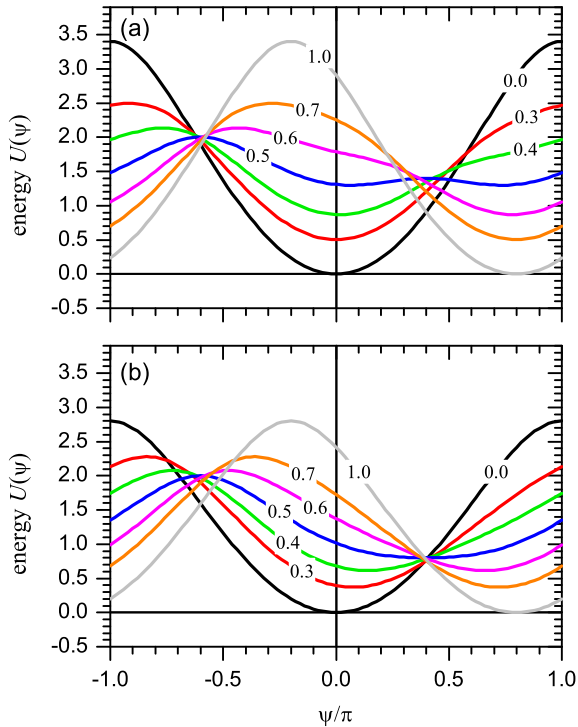}
  \caption{(Color online)
    Josephson energy $U(\psi)$ for $\bo=0.4$, $\bp=0.6$, $r_1=0.4$ and different $f=0\ldots1$ specified next to each curve.
    (a) $\alpha=0.7>|\alpha_c|=0.5$;
    (b) $\alpha=0.4<|\alpha_c|=0.5$.
  }
  \label{Fig:U(psi)@DiffPhi}
\end{figure}

Examples of $U(\psi)$ calculated for different values of applied magnetic flux are shown in Fig.~\ref{Fig:U(psi)@DiffPhi}. For $\alpha>|\alpha_c|$, see Fig.~\ref{Fig:U(psi)@DiffPhi}(a), $U(\psi)$ at $f=0$ has a single minimum at $\psi=0$. At $f=0.5$ it transforms into a double well potential. It is the same potential as one gets for $\alpha=-0.5$ and $f=0$, but shifted by (centered at) $\psi=r_1\pi=0.4\pi$ and lifted by $2\alpha=-2\alpha^\mathrm{new}=1.4$, see Eqs.~\eqref{Eq:symmetry.nPi} and \eqref{Eq:U.new}. In essence, at this $f$ one obtains a $\varphi_0\pm\varphi$ JJ. For $f=1$ the energy $U(\psi)$ transforms again to the same profile as at $f=0$, but shifted by (centered at) $\psi=r_1 2\pi=0.8\pi$, see Eqs.~\eqref{Eq:symmetry.nPi} and \eqref{Eq:U.new}. For $0<\alpha<|\alpha_c|$, see Fig.~\ref{Fig:U(psi)@DiffPhi}(b), we again start at $f=0$ from having a single minimum of $U(\psi)$ at $\psi=0$. At $f=0.5$ we again obtain the $U(\psi)$ profile as the 0-$\pi$ SQUID with $\alpha^\mathrm{new}=-\alpha$, but at $f=0$, shifted by $\psi=r_1\pi$. However, in this case it is not a double-well as $-\alpha\not<\alpha_c$. At $f=1$ we obtain the original profile (as for $f=0$), but shifted by $r_1 2\pi$, see Eqs.~\eqref{Eq:symmetry.nPi} and \eqref{Eq:U.new}.

\subsection{Ground state phase}

\begin{figure}[!tb]
  \includegraphics{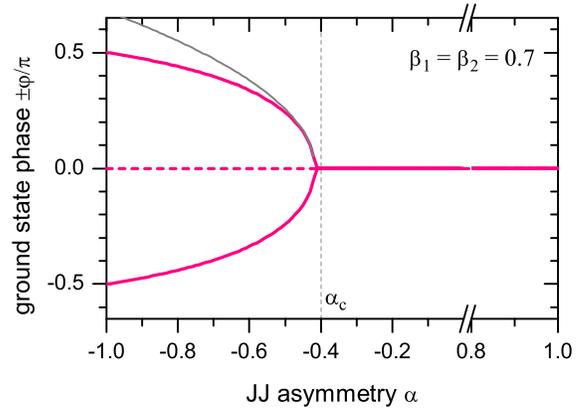}
  \caption{%
    The ground state phase $\pm\varphi(\alpha)$. The horizontal dashed line shows the unstable static solution $\psi=0$ in the region $\alpha<\alpha_c$. Thin line shows the approximation given by Eq.~\eqref{Eq:varphi_GL}.
  }
  \label{Fig:varphi(alpha)}
\end{figure}

The ground state ($f=0$, $\gamma=0$) phase $\varphi$ in the $\varphi$ domain $\alpha<\alpha_c$ can be calculated numerically from $\gamma(\varphi)=0$.

To simplify and accelerate this computation we use the following procedure. Since for $\gamma=0$ the phases $\phi_1$ and $\phi_2$ (but not $\psi$) depend only\cite{Note:phi12(beta_sum)} on $\bsum$, for calculation of $\phi_1$ and $\phi_2$, without loosing generality, we assume that $\bo=\bsum$ and $\bp=0$. Then Eqs.~\eqref{Eq:phi_12(psi)} collapse to
\begin{equation}
  \phi_2 = \phi_1+\bsum\sin(\phi_1)
  . \label{Eq:GrStt:phi_2(phi_1)}
\end{equation}
Substituting this into Eq.~\eqref{Eq:gamma} with $\gamma=0$ we arrive at a rather simple transcendental equation,
\[
  \sin(\phi_1) + \alpha\sin[\phi_1+\bsum\sin(\phi_1)]=0,
\]
which we solve to find $\phi_1$. Then the value of $\phi_2$, if necessary, can be calculated from Eq.~\eqref{Eq:GrStt:phi_2(phi_1)}. The ground state phase $\varphi$ is obtained using one of the Eqs.~\eqref{Eq:phi_12(psi)}, with $\bo$ and $\bp$ corresponding to the real circuit.

An example of $\varphi(\alpha)$ at fixed $\bboth$ is shown in Fig.~\ref{Fig:varphi(alpha)}. In essence this is $\varphi$ along a vertical line crossing a $\varphi$ domain in Fig.~\ref{Fig:VarphiDomain.3D}. The ground state phase is zero as $\alpha$ decreases from 1 down to the bifurcation point $\alpha_c$. After the bifurcation point ($\alpha<\alpha_c)$ the zero solution becomes unstable, as indicated by the dashed line. Instead two degenerate stable solutions appear.

The phase $\varphi$ corresponding to the degenerate state departs from zero as described by Eq.~\eqref{Eq:varphi_GL}.
%
%
At $\alpha\to-1$ the ground state phase tends to its maximum value $\varphi_\mathrm{max}$, which is given by
\begin{equation}
  \varphi_\mathrm{max}=\frac{\pi}{2}+ y \frac{\beta_1-\beta_2}{\beta_1+\beta_2}
  , \label{Eq:varphi_max}
\end{equation}
where $y$ is a solution of the equation
\begin{equation}
  2y = \bsum\cos(y)
  . \label{Eq:TEq4phi_2}
\end{equation}
The possible range of $y$ is from $y=0$ for $\bo=\bp=0$ to $y=y^\star$, where $y^\star\approx 0.739$ is a solution of the equation $y^\star=\cos(y^\star)$. The corresponding range of $\psi_\mathrm{max}$ is then from $\pi/2-y^\star$ (reached for $\bo\to0$, $\bp\to1$) to $\pi/2+y^\star$ (reached for $\bo\to1$ and $\bp\to0$). Note that since $y$ is a function of $\bsum$ only, it follows from \eqref{Eq:varphi_max} that the phase $\varphi_\mathrm{max}$ is antisymmetric with respect to the diagonal direction $\bo=\bp$. In particular, $\varphi_\mathrm{max}=\frac{\pi}{2}$ at $\bo=\bp$ (symmetric system). It is interesting that even for a $\pi$ JJ, which is weaker than the 0 JJ, one can obtain a ground state phase $|\varphi|>\pi$ --- a situation, which is not possible in a continuous 0-$\pi$ JJ studied earlier\cite{Goldobin:2011:0-pi:H-tunable-CPR,Sickinger:2012:varphiExp}.

\begin{figure*}[!tb]
  \includegraphics{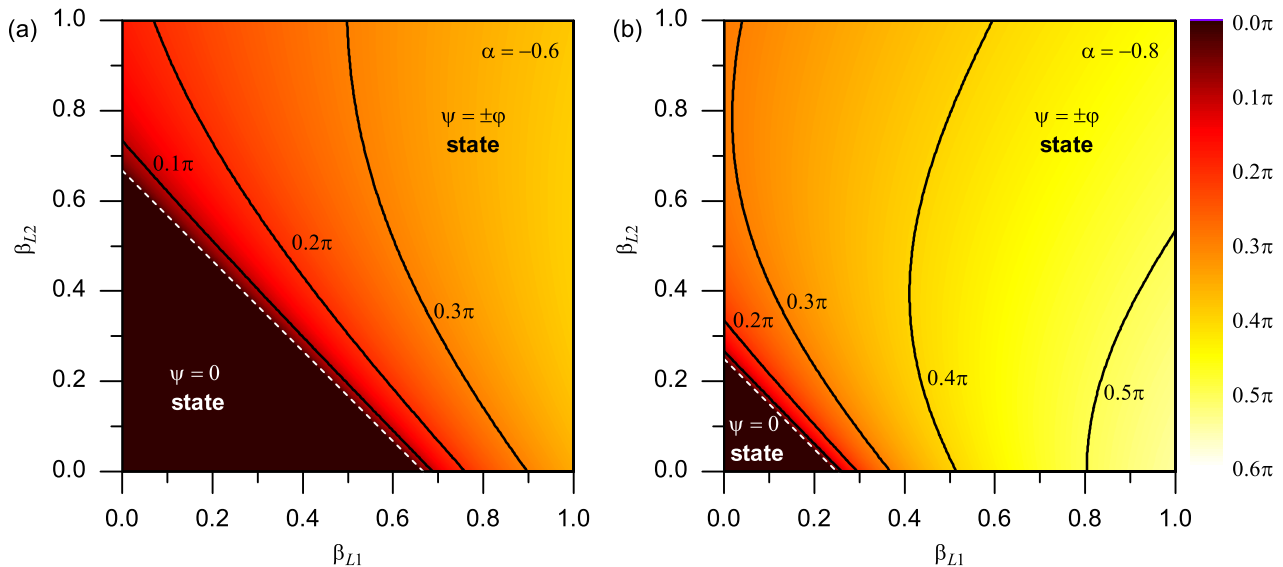}
  \caption{%
    The phase diagram of 0-$\pi$ dc SQUID (ground state phase $\varphi(\bo,\bp)$ for (a) $\alpha=-0.6$ and (b) $\alpha=-0.8$. The dashed line shows the boundary between trivial $\psi=0$ and $\psi=\pm\varphi$ ground states given by expression~\eqref{Eq:PhaseDia:beta2(beta1)}. Continuous lines are the lines of the constant ground state phase $\varphi$. Its value is given next to each line.
  }
  \label{Fig:PhaseDia}
\end{figure*}

It is also interesting to plot the ground state phase on the $(\bo,\bp)$ plane for fixed $\alpha$, \ie,  in essence, in the horizontal plane crossing the $\varphi$ domain in Fig.~\ref{Fig:VarphiDomain.3D} at fixed $\alpha$. In such a plane, with the help of Eq.~\eqref{Eq:alpha_c}, the $\varphi$ domain is given by
\begin{equation}
  \bp>\frac{-1}{\alpha}-1-\bo\text{ for } \alpha<0
  . \label{Eq:PhaseDia:beta2(beta1)}
\end{equation}
Thus, the boundary is just a straight line, see Fig.~\ref{Fig:PhaseDia}. Below this line the ground state phase is $\psi=0$, while in the filled area above this line the ground state phase is $\psi=\pm\varphi$. Note, that the boundary given by Eq.~\eqref{Eq:PhaseDia:beta2(beta1)} shifts towards the origin as $\alpha$ decreases. At $\alpha=-1$ the domain of a $0$ state vanishes completely. At $\alpha<-1$ (a case not considered here) a small $\varphi=\pi$ domain will appear close to the origin and will grow as $\alpha$ decreases further.


\subsection{Persistent current}

If the ground state of the system ($f=\gamma=0$) is the $\pm\varphi$ state, one has a persistent current circulating clockwise or counterclockwise around the SQUID. From Eq.~\eqref{Eq:gamma}, its value is given by
\begin{equation}
  I_\mathrm{circ}= \sin[\phi_1(\pm\varphi)] = \alpha\sin[\phi_2(\pm\varphi)]
  . \label{Eq:Icirc}
\end{equation}

Since in the ground state the phases $\phi_1$ and $\phi_2$ depend only\cite{Note:phi12(beta_sum)} on $\bsum$, so does $I_\mathrm{circ}$. This means that the value $\varphi$ of the ground state phase, which depends on $\bo-\bp$ and $\bsum$, can be chosen independently from the value of the persistent current $I_\mathrm{circ}$, which depends only on $\bsum$.

\begin{figure}[!htb]
  \includegraphics{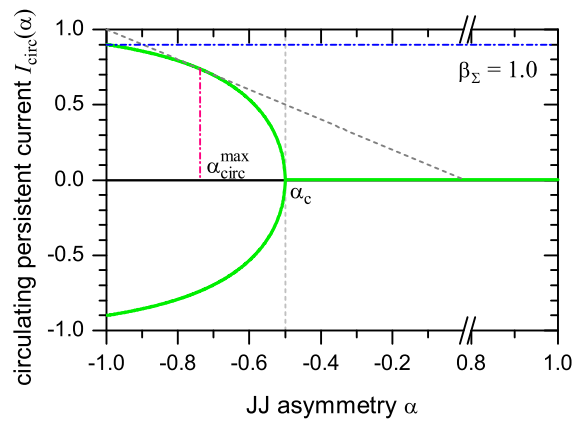}
  \caption{%
    Circulating current as a function of $\alpha$. Tilted dashed line shows the value of the critical current of the $\alpha$-junction. The horizontal dashed-dotted line shows the maximum value of persistent current reached at $\alpha=-1$ and calculated using Eq.~\eqref{Eq:phi_2.maxPersCurr}. Vertical dashed-dotted line shows the value $\alpha_\mathrm{circ}^\mathrm{max}$ calculated using Eq.~\eqref{Eq:alpha_circ^max}.
  }
  \label{Fig:CircCurr}
\end{figure}

For fixed $\bsum$ the value of persistent current grows $\propto\sqrt{\alpha_c-\alpha}$ near $\alpha_c$ and reaches the maximum value at $\alpha=-1$, see Fig.~\ref{Fig:CircCurr}. This maximum value is equal to $\sin\phi_2$, where $\phi_2$ is a solution of the following transcendental equation
\begin{equation}
  2\phi_2 - \bsum \sin\phi_2 = \pi
  . \label{Eq:phi_2.maxPersCurr}
\end{equation}
In the limit $\bsum\to0$, from Eq.~\eqref{Eq:phi_2.maxPersCurr} $\phi_2\to\pi/2$ and circulating current $\to1$.

Another interesting question is: does the value of circulating current ever equal to the maximum possible value $|\alpha|$ --- the critical current of the weaker JJ? It turns out that for fixed $\bsum$ this happens for $\alpha=\alpha_\mathrm{circ}^\mathrm{max}$, which is a solution of the following transcendental equation
\begin{equation}
  \alpha\bsum + \arcsin(\alpha) + \frac\pi2 = 0
  , \label{Eq:alpha_circ^max}
\end{equation}
see also Fig.~\ref{Fig:CircCurr}. To proof this we take the state with $\phi_2=\pi/2$, then $\sin\phi_2=1$. From Eq.~\eqref{Eq:gamma} $\sin\phi_1=-\alpha>0$, and therefore $\phi_1=\arcsin(-\alpha)$ (exactly this root!). By substituting this into Eqs.~\eqref{Eq:phi_12(psi)} and, without loosing generality assuming $\bo=0$, $\bp=\bsum$, we arrive at \eqref{Eq:alpha_circ^max}.

\subsection{Self-generated flux}

The self-generated flux in the loop (not including external flux $\phi_e$) is given by
\begin{equation}
  2\pi\frac{\Phi}{\Phi_0} = \bo\sin\phi_1(\psi)-\alpha\bp\sin\phi_2(\psi) = \phi_2(\psi)-\phi_1(\psi)-\phi_e
  . \label{Eq:Phi}
\end{equation}
Similar to the circulating current, the value of spontaneous flux in the absence of the bias ($\gamma=0$) depends only\cite{Note:phi12(beta_sum)} on $\bsum$, which is not obvious at all from the first part of Eq.~\eqref{Eq:Phi}, but apparent from its last part. This allows to write
\begin{equation}
  2\pi\frac{\Phi}{\Phi_0} = \bsum\sin\phi_1 = -\alpha\bsum\sin\phi_2 = \bsum I_\mathrm{circ}
  . \label{Eq:Phi(phi1)}
\end{equation}
Therefore, spontaneous flux behaves similar to the circulating current. The maximum value of the flux is also reached at $\alpha\to-1$.

\begin{figure}[!htb]
  \includegraphics{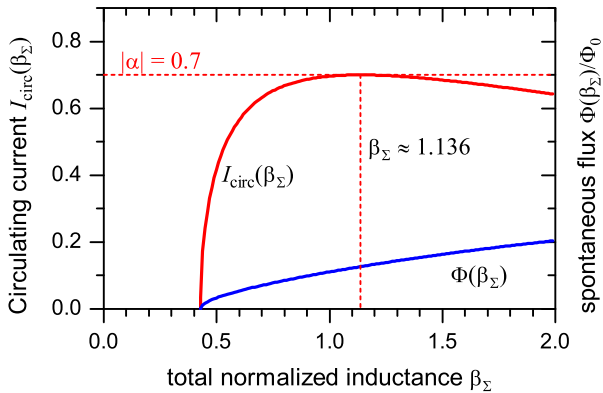}
  \caption{%
    The amplitudes of the spontaneous circulating current $I_\mathrm{circ}(\bsum)$ and spontaneous flux $\Phi(\bsum)/\Phi_0$ for fixed $\alpha=-0.7$.
  }
  \label{Fig:SpFlux&Icirc(bsum)}
\end{figure}

It is instructive to replot $I_\mathrm{circ}$ and $\Phi$ as a function of $\bsum$ at fixed $\alpha$, see Fig.~\ref{Fig:SpFlux&Icirc(bsum)}. One can see that at small inductance $\bsum<(1+\alpha)/(-\alpha)$ [inverted Eq.~\eqref{Eq:alpha_c}] the system is in the zero state. At larger inductances the system enters into $\pm\varphi$ state and the spontaneous flux and current increase. However, $\Phi(\bsum)$ grows monotonously, reaching $\sim\Phi_0/5$ at $\bsum=2$, while the $I_\mathrm{circ}(\bsum)$ exhibit a maximum, where $I_\mathrm{circ}=|\alpha|$ --- the maximum possible value in our SQUID. It happens at $\bsum=(2\arcsin(\alpha)+\pi)/(-2\alpha)$, which was obtained by inverting Eq.~\eqref{Eq:alpha_circ^max}.

\subsection{Critical currents}

In general, our system has four critical currents in the $\varphi$ domain and two critical currents outside of it. These critical currents correspond to the escape of the phase from the left ($-\varphi$) or the right ($+\varphi$) wells of the double-well potential, see Fig.~\ref{Fig:CPR}, for different directions of the bias current $\gamma$. The critical current corresponds to the maximum of the CPR $\gamma(\psi)$.

\begin{figure}[!htb]
  \includegraphics{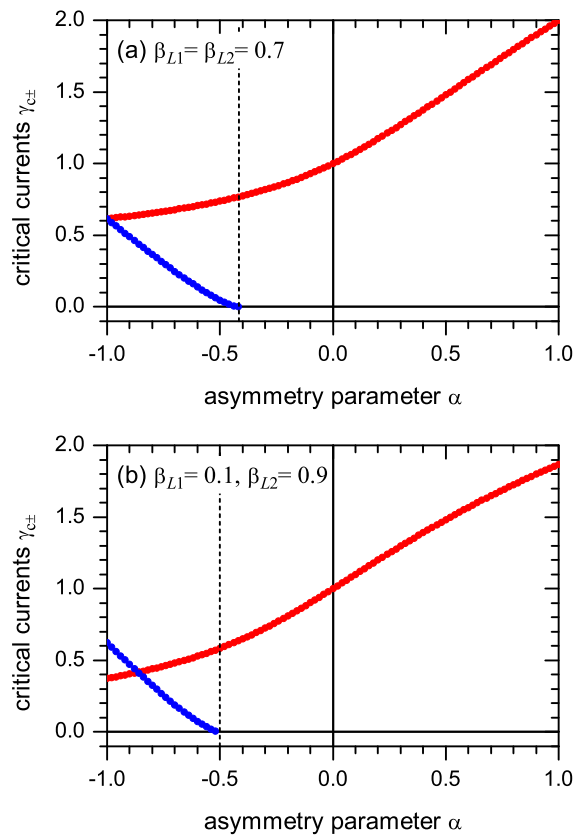}
  \caption{%
    Numerically calculated dependences $\gamma_{c\pm}(\alpha)$ for
    (a) $\bo=\bp=0.7$ and
    (b) $\bo=0.1$ and $\bp=0.9$.
  }
  \label{Fig:Ic(alpha)}
\end{figure}

The numerically calculated dependence of $\gamma_{c}(\alpha)$ for given $\bo$ and $\bp$ is shown in Fig.~\ref{Fig:Ic(alpha)}. One can see several key points on the $\gamma_{c\pm}(\alpha)$ dependence. First, $\gamma_{c+}(0)\equiv1$. Second, for $\bo=\bp$, $\gamma_{c-}(-1)=\gamma_{c+}(-1)$, \ie, the two dependences converge at $\alpha=-1$, see Fig.~\ref{Fig:Ic(alpha)}(a). In the case $\bo<\bp$, see Fig.~\ref{Fig:Ic(alpha)}(b), the dependences cross at some $-1<\alpha_x<\alpha_c$.

\subsection{Critical current as a function of applied flux}

\begin{figure}[!htb]
  \includegraphics{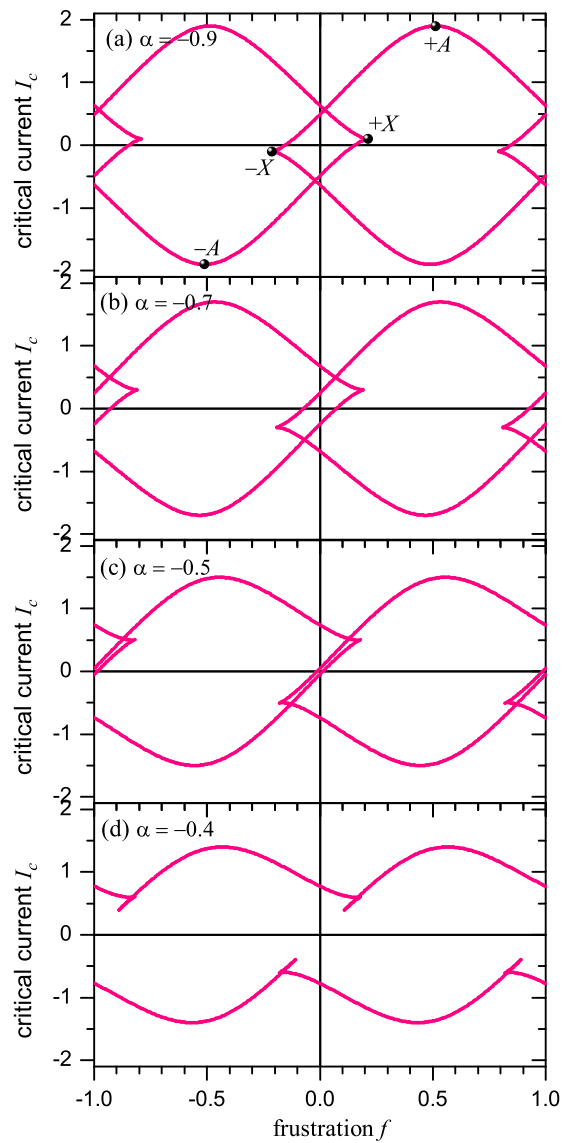}
  \caption{%
    The dependence of the critical current $\gamma_c$ of the device on the normalized applied magnetic field (frustration) $f$ for $\bo=\bp=0.7$ ($\pi\beta_L=1.4$, $\alpha_c\approx -0.417$) and different $\alpha$.
  }
  \label{Fig:Ic(H)}
\end{figure}

The presence of the degenerate ground state and two critical currents also manifests itself in the dependence of the maximum supercurrent on magnetic field $\gamma_c(f)$. In the general case this dependence can be calculated only numerically.
Several examples of $\gamma_c(f)$ dependences are presented in Fig.~\ref{Fig:Ic(H)}. For the parameters in the $\varphi$ domain ($\alpha<\alpha_c$) one sees the domains corresponding to the different flux states overlapping in the vicinity of integer $f$. In this overlapping region one observes in total four critical currents corresponding to the escape of the phase from different energy minima (wells) of the Josephson potential in two different directions. Note the striking similarity of these curves with those for $\varphi$ JJs based on continuous 0-$\pi$ JJs\cite{Goldobin:2011:0-pi:H-tunable-CPR}.

As $\alpha$ increases and approaches $\alpha_c$, see Fig.~\ref{Fig:Ic(H)}, the bistability region is shrinking. Exactly at $\alpha=\alpha_c$ the overlap near integer $f$ disappears, which corresponds to the disappearance of two distinct ground states $\psi=\pm\varphi$. However, for a small range of $\alpha>\alpha_c$, one observes a small triangular-like bistability region where the branches meet, see Fig.~\ref{Fig:Ic(H)}(d). Further investigation of this small domain is outside the scope of this paper. We note that the intersecting domains, similar to those shown in Fig.~\ref{Fig:Ic(H)} were calculated long ago\cite{Fulton:1972:dc-SQUID.Ic(H),Peterson:1979:SQUID.Ic(H)}. At that time, however, $\pi$ junctions were unknown, so only a 0-0 SQUID, where the domains intersect near half-integer $f$, was considered. However some key results can be adopted to our case easily.

In particular, we can find the positions of key points $\pm A$ and $\pm X$ where $\partial\gamma_c(f)/\partial f=0$. We note\cite{Fulton:1972:dc-SQUID.Ic(H),Peterson:1979:SQUID.Ic(H)} that the phases $\phi_1=\pm\pi/2$ and $\phi_2=\pm\pi/2$ (in any combination) satisfy the equation $\gamma'=0$ for \emph{any} asymmetry and applied flux. It turns out\cite{Fulton:1972:dc-SQUID.Ic(H)} that each of these four combinations corresponds to an extremum of the $\gamma(f)$ dependence, \ie, to points $\pm A$ and $\pm X$ in Fig.~\ref{Fig:Ic(H)}. To find the values of the external flux $2\pi f\equiv\phi_e$,   corresponding to these points, we substitute the above phases into Eq.~\eqref{Eq:phi_12(psi)} and obtain
\begin{equation}
  \phi_e = \pm \left[ \frac{\pi}{2} + \bo \right] \mp \left[ \frac{\pi}{2}+\alpha\bp \right]
  . \label{Eq:phi_e.extr}
\end{equation}
The corresponding values of the critical current are obtained from Eq.~\eqref{Eq:gamma}
\begin{equation}
  \gamma_c^\mathrm{extr} = \pm 1 \pm \alpha
  . \label{Eq:gamma_c.extr}
\end{equation}
In Eqs.~\eqref{Eq:phi_e.extr} and \eqref{Eq:gamma_c.extr} both $\pm$ signs are independent, providing four combinations in total. The trivial consequence of Eq.~\eqref{Eq:gamma_c.extr} is that for a SQUID with symmetric critical currents ($|\alpha|=1$) the  points $\pm X$ are situated at the horizontal axis ($\gamma_c=0$).

The bistability region around integer $f$ can be used to store one bit of information in the $\pm\varphi$ states as demonstrated recently\cite{Goldobin:2013:varphi-bit}. In some sense it is similar to the earlier proposals\cite{Zappe:1974:SFQ-bit,Gueret:1977:SFQ-bit} to use $n=0$ and $n=1$ states of the SQUID biased to $f\approx1/2$. However in our case, the flux bias is not needed. Asymmetry also provides different critical currents at $f=0$, which simplifies readout.

Finally, the practical question is: can one detect $\pm\varphi$, $\varphi_0$ and $\varphi_0\pm\varphi$ states in experiment by measuring the $\gamma_c(f)$ dependence? For the $\pm\varphi$ state the answer is given in Fig.~\ref{Fig:Ic(H)}. One should observe intersection of branches and four critical currents. For the $\varphi_0\pm\varphi$ state the situation is similar. However, since this state appears at half-integer $f$ the whole $\gamma_c(f)$ curve is shifted so that the bistability regions are situated around half-integer $f$. Finally, in the $\varphi_0$ state, which in our system appears at finite field only, one has only two critical currents and the junction looks just like a conventional one, although the $\gamma_c(f)$ dependence is unusual (periodic), but never multi-state. To prove the $\varphi_0$ state one has to do a phase sensitive experiment, \eg, putting our black box in a superconducting loop.

\section{Summary}
\label{Sec:Conclusions}

We have shown that an asymmetric 0-$\pi$ SQUID can be used as an effective $\varphi$ JJ with magnetic field tunable current-phase relation $\gamma(\psi)$ and, accordingly, a Josephson energy $U(\psi)$. The critical value $\alpha_c$ of the critical current asymmetry parameter $\alpha$ required to obtain the degenerate $\pm\varphi$ ground state depends on the sum $L_1+L_2$ of inductances in two branches of the SQUID. Upon applying an integer number of flux quanta $f=n$, the phase $\psi$ across the structure advances by an amount $r_1\cdot2\pi n$, where $r_1$ is the fraction of external flux induced in the left branch of the SQUID. Since, in general, $r_1$ is an arbitrary number depending on design, the phase shift $r_1\cdot2\pi n$ is not a multiple of $2\pi$. By applying a half-integer number of flux quanta $f=n+1/2$ to a 0-0 SQUID with $\alpha<|\alpha_c|$, one can turn the effective 0-JJ into a $\varphi_0=r_1 \pi n$ JJ. If $\alpha>|\alpha_c|$, then one can turn the effective 0-JJ into a JJ with ground states $\varphi_0\pm\varphi$. The dependence of the critical current on magnetic flux clearly shows bistability regions typical for $\varphi$ JJ\cite{Goldobin:2011:0-pi:H-tunable-CPR,Sickinger:2012:varphiExp}.

In terms of designing a practical device (bistable $\varphi$ JJ) the target parameters can be, \eg, $\bo=\bp=0.4\ldots0.7$ and $\alpha\approx-0.7\ldots-0.8$ to stay well inside the $\varphi$ domain in Fig.~\ref{Fig:VarphiDomain.3D}. This will provide very large operation margins. Note that a finite inductance is essential to obtain a $\varphi$ domain of finite size. In the limit $\bsum\to0$ the $\varphi$-domain shrinks to a point.

\acknowledgments

We thank M. Fistul, V. Ryazanov, A. Ustinov, M. Weides for useful discussions. This work was funded by the Deutsche Forschungsgemeischaft (DFG) via projects No. GO-1106/5 and No. SFB/TRR 21 A5 and by the EU-FP6-COST Action MP1201.

\appendix

\section{Solutions close to the bifurcation point $\alpha\approx\alpha_c$}
\label{Sec:GL}

The $\varphi$ JJ proposed here can also be used as a qubit at $f=0$, when the barrier separating the $-\varphi$ and $+\varphi$ states is small. This is the case when $\alpha$ is only slightly smaller than $\alpha_c$. In this limit the expressions for many important quantities can be obtained analytically. We are especially interested in the situation near the bottom of the energy profile $U(\psi)$, \ie, for small values of $\psi$ since the important physics (formation of $\pm\varphi$ state, escape, macroscopic quantum tunneling) takes place there.

For $\alpha=\alpha_c-\epsilon$ ($0\leq\epsilon\ll1$) the Josephson energy of the system can be expanded like in the Ginzburg-Landau theory as
\begin{equation}
  U_\mathrm{GL}(\psi) \approx a \psi^2 + b \psi^4,\quad |\psi|\ll 1
  . \label{Eq:U_GL}
\end{equation}
where the coefficients $a(\alpha)$ and $b(\alpha)$ have to be determined from our model given by Eqs.~\eqref{Eq:phi_12(psi)} and \eqref{Eq:gamma}, namely, from
\begin{eqnarray}
  U''(0) &=& \gamma'(0) = 2a
  ; \label{Eq:a.gen}\\
  U''''(0) &=& \gamma'''(0) = 24 b
  , \label{Eq:b.gen}
\end{eqnarray}
From Eq.~\eqref{Eq:gamma}
\begin{equation}
  \gamma'(\psi) = \cos(\phi_1)\phi'_1 + \alpha \cos(\phi_2) \phi_2'
  . \label{Eq:gamma'}
\end{equation}
The derivatives $\phi'_{1,2}$ can be calculated by differentiating Eqs.~\eqref{Eq:phi_12(psi)} with respect to $\psi$:
\begin{equation}
  \phi_1' = \frac{1}{1+\bo      \cos(\phi_1)}, \quad
  \phi_2' = \frac{1}{1+\bp\alpha\cos(\phi_2)}
  . \label{Eq:phi12'}
\end{equation}
By substituting $\phi'_{1,2}$ from Eq.~\eqref{Eq:phi12'} into Eq.~\eqref{Eq:gamma'} we obtain
\begin{equation}
  \gamma'(\psi)=\frac{\cos(\phi_1)}{1+\bo      \cos(\phi_1)} + \frac{\alpha\cos(\phi_2)}{1+\bp\alpha\cos(\phi_2)}
  . \label{Eq:gamma'.final}
\end{equation}
According to Eqs.~\eqref{Eq:phi_12(psi)}, $\psi=0$ corresponds to $\phi_1=\phi_2=0$, so from Eqs.~\eqref{Eq:a.gen} and \eqref{Eq:gamma'.final} we obtain the explicit expression for $a(\alpha)$:
\begin{equation}
  a(\alpha) = \frac12 \left[ \frac{1}{1+\bo} + \frac{\alpha}{1+\alpha\bp}\right]
  , \label{Eq:a.final}
\end{equation}
which is negative for $\alpha<\alpha_c$. Near $\alpha_c$ the leading term is
\begin{equation}
  a(\epsilon) = -\frac12 \frac1{\alpha_c^2 (1+\bo)^2} \epsilon = a_1 \epsilon
  . \label{Eq:a1}
\end{equation}

Similarly, differentiating $\gamma'(\psi)$ in Eq.~\eqref{Eq:gamma'.final} two additional times and using
Eq.~\eqref{Eq:phi12'} after each differentiation, we obtain at $\psi=\phi_1=\phi_2=0$ from Eq.~\eqref{Eq:b.gen}
\begin{equation}
  b(\alpha) = \frac{-1}{24} \left[ \frac{1}{(1+\bo)^4} + \frac{\alpha}{(1+\alpha\bp)^4}\right]
  . \label{Eq:b.final}
\end{equation}
Near $\alpha_c$ the main term of $b(\alpha)$ is a constant
\begin{equation}
  b(\alpha_c) = \frac{1}{24} \frac{\bsum (\bsum^2+3\bsum+3)}{(1+\bo)^4}
  = \frac1{24} \frac{\alpha_c^3+1}{(-\alpha_c)^3 (1+\bo)^4} = b_0
  . \label{Eq:b0}
\end{equation}

Now we can readily calculate various quantities. The value $\varphi$ of the ground state phase is determined from $\gamma_\mathrm{GL}(\psi)=U'_\mathrm{GL}(\psi)=0$, \ie,
\begin{equation}
  \varphi_\mathrm{GL} = \sqrt{\frac{-a}{2b}}
  \approx \sqrt\frac{-a_1}{2b_0}\sqrt{\epsilon}
  . \label{Eq:varphi_GL}
\end{equation}

The self-generated flux in the ground state is given by Eq.~\eqref{Eq:Phi} and for $\alpha\to\alpha_c$ can be approximated by
\begin{equation}
  2\pi\frac{\Phi_\mathrm{GL}}{\Phi_0}\approx \frac{\bsum}{1+\bo}\varphi_\mathrm{GL}
  =\frac{\bsum}{1+\bo}\sqrt\frac{-a_1}{2b_0}\sqrt{\epsilon}
  . \label{Eq:Phi.approx}
\end{equation}
where we took into account that for small $\psi$, according to Eq.~\eqref{Eq:phi_12(psi)}, $\phi_1\approx\psi/(1+\bo)$ and $\phi_2\approx\psi/(1+\alpha\bp)$.

The barrier between two wells is given by
\begin{equation}
  U_\mathrm{GL}(0)-U_\mathrm{GL}(\varphi) = \frac{a^2}{2b} \approx \frac{a_1^2}{2b_0} \epsilon^2
  . \label{Eq:DeltaU_GL}
\end{equation}

The depinning current $\gamma_{c-}$ can be found as an extremum of $\gamma(\psi)$. The extremum is reached at $\psi=\psi_\mathrm{dep}$ satisfying the equation $\gamma'(\psi_\mathrm{dep})=2a+12b\psi_\mathrm{dep}^2=0$. Thus,
\begin{equation}
  \psi_\mathrm{dep} = \sqrt{\frac{-a}{6b}} = \frac{\varphi_\mathrm{GL}}{\sqrt3}
  \approx \sqrt{\frac{-a_1}{6b_0}}\sqrt{\epsilon}
  . \label{Eq:psi_dep}
\end{equation}
The value of $\gamma_{c-}=\gamma(\psi_\mathrm{dep})$ is
\begin{equation}
  \gamma_- = \psi_\mathrm{dep}\frac43 a = 4\sqrt{\frac{-a_1^3}{6b_0}} \epsilon^\frac32
  . \label{Eq:gamma_-.GL}
\end{equation}
The value of $\gamma_{c+}$ cannot be calculated in the framework of our GL-approximation as $\gamma_{c+}$ corresponds to a large depinning phase.

The eigenfrequency in each of the $\pm\varphi$ wells can be calculated as
\begin{equation}
  \omega_0=U''(\varphi)=\gamma'(\varphi) = -4a
  \approx -4a_1\epsilon
  . \label{Eq:omega_0.GL}
\end{equation}


\clearpage
\bibliography{this,JJ,pi,SF,SFS,MyJJ}

\begin{thebibliography}{38}
\bibitem [{Not()}]{Note:phi12(beta_sum)}%
  \BibitemOpen
  \href@noop {} {}\bibinfo {note} {The values of phases $\phi_1$ and $\phi_2$
  in the absence of applied current ($\gamma=0$) depend only on $\bsum$ (or,
  equivalently, on $\alpha_c$). Physically, this is obvious since on the
  equivalent circuit the inductances are in series for $\gamma=0$. To show this
  mathemetically, we assume that we have some static solution $\phi_1$ and
  $\phi_2$ of Eqs.~\eqref{Eq:phi_12(psi)} and \eqref{Eq:gamma} for given $\bo$
  and $\bp$. Let us modify $\bo$ and $\bp$ as follows: $\bo\to\bo+\Delta$ and
  $\bp\to\bp-\Delta$, so that $\bsum$ remains constant. We are going to show
  that the solutions $\phi_1$, $\phi_2$ remain unchanged. By substituting the
  new $\bo$ and $\bp$ into Eqs.~\eqref{Eq:phi_12(psi)} and excluding $\psi$, we
  see that the resulting equation is satisfied for any $\Delta$ only if
  $\sin(\phi_1)=-\alpha\sin(\phi_2)$. The latter is indeed true for $\gamma=0$,
  see Eq.~\eqref{Eq:gamma}.}\BibitemShut {Stop}%
\end{thebibliography}

\end{document}